# Simulation-in-the-Reasoning (SiR): A Conceptual Framework for Empirically Grounded AI in Autonomous Transportation

Wuping Xin [1][0000-0002-9021-5763]

[1] Caliper Corporation, Newton, MA 02461, USA
`wuping@caliper.com`

**Abstract.** Large Language Models (LLMs) have advanced reasoning through techniques like Chain-of-Thought (CoT). However, their reasoning largely remains textual and hypothetical, lacking empirical grounding in complex, dynamic domains like transportation. This paper introduces **Simulation-in-the-Reasoning (SiR)**, a novel conceptual framework that embeds domain-specific simulators directly into the LLM reasoning loop. By treating intermediate reasoning steps as **executable simulation experiments**, SiR transforms LLM reasoning from narrative plausibility into a falsifiable, hypothesis-simulate-analyze workflow. We discuss applications, where LLM can formulate Intelligent Transport System (ITS) strategy hypotheses, invoke a traffic simulator via the Model Context Protocol (MCP), evaluate results under different demand patterns, and refine strategies through verification and aggregation. While implementing the framework is part of our ongoing work, this paper primarily establishes the conceptual foundation, discusses design considerations like API granularity, and outlines the vision of SiR as a cornerstone for interactive transportation digital twins. We argue that SiR represents a critical step towards trustworthy, empirically-validated AI for autonomous transportation systems.

## 1 Introduction

Large Language Models (LLMs) have shown promising reasoning abilities through methods such as Chain-of-Thought (CoT) prompting [1], self-consistency sampling [2], reinforcement learning with verifiers [3], and retrieval-augmented reasoning [4]. While these approaches improve accuracy on tasks with uniquely verifiable answers, they remain confined to **text-based reasoning**. The intermediate steps they generate are narrative hypotheses rather than empirically tested, which limit their reliability in complex real-world domains such as traffic and transportation systems.

This paper introduces **Simulation-in-the-Reasoning (SiR)**, a new framework in which LLMs embed domain-specific simulators directly into their reasoning loop. SiR extends the CoT paradigm from "step 1, step 2, step 3" in text to a **hypothesis–simulate–analyze** process. Treating simulations as executable reasoning steps, SiR grounds LLM outputs in verifiable experiments.

The contributions of this paper:

- **Conceptual Framework**: We introduce and formalize SiR as a reasoning framework, where LLMs embed domain-specific simulators directly into their reasoning loop. This elevates reasoning from a text-only narrative to an executable empirical process, allowing reasoning traces to be validated through simulation.
- **Technical Implementation**: We present the use of the Model Context Protocol (MCP) to couple an LLM's reasoning process with a professional-grade traffic simulator. This mitigates the gap between plausible-sounding intermediate steps and empirically validated outcomes, thereby an additional layer of reliability.
- **Practical Application**: We apply this framework to traffic signal optimization. In this application, LLMs generate hypotheses about signal strategies, invoke simulations through MCP, and iteratively refine strategies by verifying and aggregating results. This illustrates how SiR can be used to address **complex and dynamic problems** where text-based reasoning may be insufficient.

## 2 Background

### 2.1 Chain-of-Thought (CoT)

Reasoning in Large Language Models (LLMs) began with a simple question: *can we make a model think step by step*? Earlier experiments showed that if a prompt included phrases like "*let's think step by step*", the model suddenly solved problems it previously failed [1]. This effect, now known as Chain-of-Thought (CoT) prompting, revealed that LLMs could benefit from producing intermediate reasoning tokens. Subsequent formal analysis [2] confirmed that inherently serial problems, where later steps depend on earlier ones, are computationally intractable for a standard transformer model unless it articulates these intermediate steps. This means CoT is not just a "prompting trick", but a necessary mechanism that allows LLMs to overcome their architectural limitations and handle problems requiring sequential logic and causal inference.

### 2.2 Supervised Finetuning (SFT)

CoT on its own depends on how the prompt was phrased, and its generated steps were words that are plausible rather than testable. Researchers tried to address this through Supervised Finetuning (SFT), collecting datasets where humans wrote out problems and detailed solutions [3, 4]. It was hoped that by imitating expert reasoning, LLMs would generalize. Disappointingly, scaling SFT lead to diminishing returns: the models got better at familiar patterns but often stumbled when faced with new problem structures [4, 5].

### 2.3 Reinforcement Learning with Verifiers

A natural next step was to let LLMs learn from their **own reasoning traces**. This gave rise to **reinforcement learning with verifiers**. Here, the LLM generates candidate solutions, and an automated verifier decides whether they are correct [5,6]. Importantly, this line of work showed that the success of RL did not hinge on optimization but on the quality of the verifier. In other words, *verification is the key to reasoning at scale*. This resonates strongly with simulation: traffic simulators, for example, inherently enforce physical constraints like flow conservation and queue spillback, serving as powerful verifiers embedded in the environment.

Even with verifiers, models remained brittle. In CoT, a path represents one complete reasoning trajectory the LLM generates, i.e., a sequence of steps leading to an answer. Different decoding runs (e.g., with sampling or beam search) can produce different reasoning paths, among which there are the so-called weaker paths, ones that look fluent or confident but are logically flawed or yield an incorrect answer. Such weaker paths often compete with the correct one, and sometimes the decoding methods selected the wrong outcome.

The fix came with **self-consistency** [7]. Instead of trusting a single reasoning trajectory, sample many, then choose the most common answer. This simple mechanism, essentially majority voting across reasoning paths, improved accuracy dramatically. When extended to simulation, the analogy is straightforward: run the same strategy under multiple stochastic seeds or demand scenarios, then trust the result that consistently holds up.

### 2.4 Model Context Protocol as the Glue

Across all these advances, a common thread remains: the **reasoning steps themselves are textual**. They sound convincing, but are not tied to physical reality. This is the gap that motivates the work in this paper.

We argue that the next logical extension of CoT is to let intermediate steps be **experiments**—to embed simulators inside the reasoning loop. The **Model Context Protocol (MCP),** introduced by Anthropic in 2024 [9], is an open standard that makes this technically feasible [10].

MCP provides a structured interface for LLMs to discover and invoke external tools and data sources, exposing task-specific APIs via MCP servers and letting the LLM act as an MCP client [9, 10]. Through MCP, an LLM can construct a traffic scenario, run a microscopic simulator such as TransModeler, retrieve performance measures, and incorporate them back into its reasoning. This transforms reasoning from narra-tive plausibility into **empirically grounded logic**, closing the gap between language and the physical systems it seeks to describe.

In this paper, we introduce and formalize the **Simulation-in-the-Reasoning (SiR)** framework, which leverages MCP as the integration layer between LLMs and simulators. MCP elevates simulation calls into the **core reasoning process**. In SiR, an LLM generates a hypothesis, invokes a simulator such as TransModeler through MCP, retrieves empirical measures (e.g., delays, throughput, emissions), and then updates its reasoning trajectory based on these outcomes. This closes the loop between language-based reasoning and experimentally grounded validation. By combining CoT with MCP-enabled simulation, SiR provides a principled pathway to move from *narrative plausibility* toward scientifically grounded AI reasoning in transportation and beyond.

## 3  The SiR Framework

The **Simulation-in-the-Reasoning (SiR)** framework extends the Chain-of-Thought paradigm by embedding domain-specific simulators into the reasoning loop of Large Language Models (LLMs). The core principle is to transform intermediate reasoning steps from purely textual hypotheses into **executable experiments**. This section formalizes the main components of SiR, describes the reasoning workflow, and presents a conceptual diagram of the framework.

### 3.1  Components

There are three components in the SiR framework: the **LLM Agent**, the **Simulator**, and the **Model Context Protocol (MCP) Interface**. Together, they define how reasoning is generated, tested, and grounded in empirical reality.

The **LLM Agent** serves as the central reasoning engine. It decomposes complex transportation problems into structured steps and generates hypotheses in natural language. Beyond hypothesis generation, the LLM must decide when to invoke external tools, ensuring that simulations are executed only when they add value. Once results are returned, the LLM interprets these outputs, evaluates them against objectives such as minimizing delay or preventing spillback, and iteratively refines strategies through repeated reasoning–simulation cycles.

The **Simulator** provides an empirical grounding mechanism. Unlike text-based reasoning, which risks remaining hypothetical, the simulator executes experiments based on hypotheses proposed by LLM. In transportation, this corresponds to a traffic simulator such as *TransModeler* [11], which models vehicle dynamics, traffic signal operations, and complex demand patterns. The simulator produces quantitative performance measures—such as travel times, throughput, queue lengths, and emissions—that serve as verifiers of the LLM's reasoning.

Finally, the **Model Context Protocol (MCP) Interface** acts as the integration layer between the LLM and the simulator. MCP exposes simulator functions as structured APIs that the LLM can discover and invoke in a standardized way. Through MCP, the LLM passes scenario specifications, initiates simulation runs, and parses structured results back into its reasoning process. Ensuring reproducibility and portability, MCP elevates simulation calls from peripheral tool use to integral reasoning steps.

Table 1. Components of the SiR framework.

| Component | Role | Key Functions |
|---|---|---|
| LLM Agent | Central reasoning engine | Generate hypotheses, decompose problems, decide on tool use, interpret results, refine strategies |
| Simulator | Empirical grounding mechanism | Execute experiments, model traffic systems, output performance measures (delay, queues, emissions) |
| MCP Interface | Integration layer (LLM ↔ Simulator) | Expose APIs, pass scenario specifications, invoke runs, parse structured results, ensure reproducibility |

## 3.2 Workflow

The SiR framework structures reasoning as an iterative loop of hypothesis–simulate–analyze–refine (See Figure 1).

It begins with problem formulation, where the LLM interprets a task (e.g., signal optimization) and defines objectives such as minimizing delay or avoiding spillback. In hypothesis generation, the model proposes candidate strategies using Chain-of-Thought reasoning, such as adjusting cycle length or phasing.

Through simulation invocation, the LLM uses the Model Context Protocol (MCP) to configure and call a microscopic simulator. The simulator executes one or more runs, possibly under varied random seeds to capture stochastic dynamics.

The outputs—result parsing—are returned as structured performance measures (e.g., delay, throughput, queues). The LLM evaluates these results against constraints and objectives.

Finally, in analysis and refinement, the model updates its hypotheses and repeats the cycle until a robust solution emerges. This workflow transforms reasoning from text-only speculation into empirically grounded analysis.

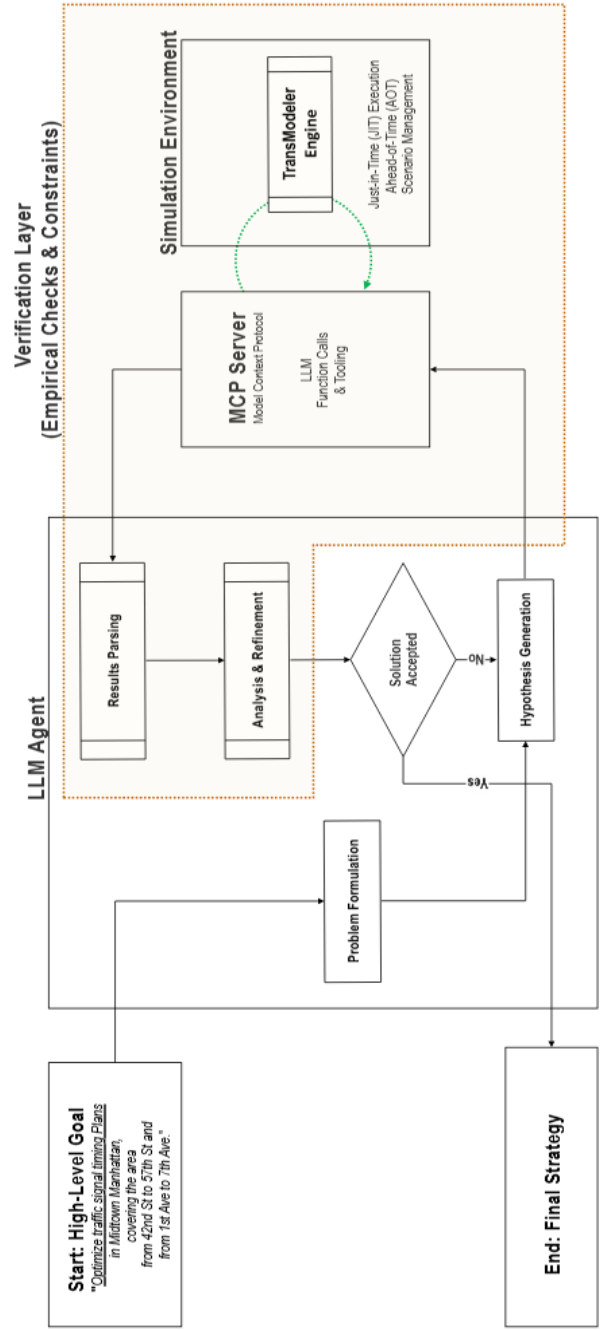

**Fig. 1**. SiR workflow diagram.

# 4 Discussion: Implications and Design Considerations

## 4.1 MCP Nuances

The Model Context Protocol (MCP) plays a central role in making the Simulation-in-the-Reasoning (SiR) framework technically feasible. MCP defines how large language models discover, invoke, and parse simulator functions, transforming what would otherwise be ad hoc tool calls into structured reasoning steps. For SiR, this standardization ensures reproducibility across different simulators and domains while isolating the LLM from low-level API complexity. In practice, this means that the reasoning loop can be portable, transparent, and systematically verifiable, rather than tied to the idiosyncrasies of a single software tool.

## 4.2 API Granularity

A critical design issue for SiR is the granularity of simulator APIs exposed to MCP. If the APIs are too coarse-grained—for example, offering only an "optimize corridor" command—the LLM's reasoning risks becoming opaque, with limited control or visibility into the intermediate steps. On the other hand, if the APIs are too fine-grained—such as setting every cycle length, phase split, or retrieving queue lengths at specific nodes—the complexity may overwhelm the reasoning agent and demand excessively detailed prompt engineering. The success of SiR therefore hinges on carefully balancing expressiveness and simplicity in simulator API design, enabling LLMs to formulate hypotheses that are meaningful and tractable without drowning in unnecessary detail.

The strengths of SiR lie in its capacity to transform LLM outputs into empirically grounded reasoning. By running hypotheses through simulation, the framework mitigates the risk of hallucination and anchors conclusions in measurable outcomes. The process produces falsifiable and reproducible workflows, qualities that are essential for scientific credibility. More importantly, SiR makes it possible to address domains characterized by dynamic interacting processes—such as traffic and transportation systems—where purely text-based reasoning often fails to capture the complexity of reality.

## 4.3 Scalability

At the same time, the framework introduces significant challenges. One challenge is the complexity of prompt and control design: the LLM must frame hypotheses at the right level of abstraction to meaningfully interact with the simulator. Another is computational cost, since multiple stochastic simulation runs are often required within each reasoning cycle to achieve reliable conclusions. Finally, scalability remains a hurdle. While optimizing a single intersection may be tractable, extending the SiR framework to city-scale networks will require hierarchical reasoning strategies to manage complexity.

### 4.4 Digital Twin Vision

Looking beyond these near-term considerations, SiR also points toward a broader **digital twin vision** for transportation. By embedding reasoning loops directly into the twin, such a system could generate hypotheses, run simulations, and propose interventions in real time. This shift moves from simply *modeling reality* to actively *reasoning with reality*, positioning SiR as a foundation for adaptive transportation AI.

## 5 Conclusion

This paper introduced **Simulation-in-the-Reasoning (SiR)**, a framework that extends the reasoning capabilities of large language models by embedding simulators directly into the reasoning loop. The framework leverages the Model Context Protocol (MCP) to connect LLMs with domain-specific simulators. While challenges remain in areas such as API design, computational efficiency, and scalability, SiR provides a principled path toward robust, scientifically grounded reasoning. More broadly, it opens the door to interactive digital twins that move beyond passive monitoring to active reasoning and intervention, laying the foundation for a new generation of intelligent, adaptive transportation systems.